\documentclass[journal=apchd5,manuscript=article]{achemso}

\usepackage[version=3]{mhchem} 
\usepackage[T1]{fontenc}       
\usepackage{soul,color}

\author{Sebastian K. H. Andersen}
\email{sekh@mci.sdu.dk}
\affiliation[SDU University]
{Center for Nano Optics, University of Southern Denmark, DK-5230 Odense M, Denmark}

\author{Simeon Bogdanov}
\affiliation[Birck Nanotech]
{School of Electrical and Computer Engineering, Purdue Quantum Center, Purdue University, West Lafayette, Indiana 47907, USA}

\author{Oksana Makarova}
\affiliation[Birck Nanotech]
{School of Electrical and Computer Engineering, Purdue Quantum Center, Purdue University, West Lafayette, Indiana 47907, USA}

\author{Yi Xuan}
\affiliation[Birck Nanotech]
{School of Electrical and Computer Engineering, Purdue Quantum Center, Purdue University, West Lafayette, Indiana 47907, USA}

\author{Mikhail Y. Shalaginov}
\affiliation[Birck Nanotech]
{School of Electrical and Computer Engineering, Purdue Quantum Center, Purdue University, West Lafayette, Indiana 47907, USA}

\author{Alexandra Boltasseva}
\affiliation[Birck Nanotech]
{School of Electrical and Computer Engineering, Purdue Quantum Center, Purdue University, West Lafayette, Indiana 47907, USA}

\author{Sergey I. Bozhevolnyi}
\affiliation[SDU University]
{Center for Nano Optics, University of Southern Denmark, DK-5230 Odense M, Denmark}

\author{Vladimir M. Shalaev}
\affiliation[Birck Nanotech]
{School of Electrical and Computer Engineering, Purdue Quantum Center, Purdue University, West Lafayette, Indiana 47907, USA}

\title[An \textsf{achemso} demo]
  {Hybrid Plasmonic Bullseye Antennas for Efficient Photon Collection}

\abbreviations{QE: Quantum Emitter; TiO$_2$: Titania; SiO$_2$: Silica; SI: Supporting Infomation ; NV-center: Nitrogen vacancy center ; ND: Nanodiamond ; BFP: Back focal plane; AFM: Atomic force microscope ; ZPL: Zero phonon line, SPP: Surface plasmon polariton}

\begin{document}

\setstcolor{red}			







\begin{abstract}
We propose highly efficient hybrid plasmonic bullseye antennas for collecting photon emission from nm-sized quantum emitters. In our approach, the emitter radiation is coupled to surface plasmon polaritons that are consequently converted into highly directional out-of-plane emission. The proposed configuration consists of a high-index titania bullseye grating separated from a planar silver film by a thin low-index silica spacer layer. Such hybrid systems are theoretically capable of directing 85\% of the dipole emission into a 0.9 NA objective, while featuring a spectrally narrow-band tunable decay rate enhancement of close to 20 at the design wavelength. Hybrid antenna structures were fabricated by standard electron-beam lithography without the use of lossy adhesion layers that might be detrimental to antenna performance. The fabricated antennas remained undamaged at saturation laser powers exhibiting stable operation. For experimental characterization of the antenna properties, a fluorescent nanodiamond containing multiple nitrogen vacancy centers (NV-center) was deterministically placed in the bullseye center, using an atomic force microscope. Probing the NV-center fluorescence we demonstrate resonantly enhanced, highly directional emission at the design wavelength of 670\,nm, whose characteristics are in excellent agreement with our numerical simulations.
\end{abstract}

\section{Introduction}
Efficient collection of photons from single quantum emitters (QE) is a key requirement for many quantum technological applications\cite{QuantumTechnology}, utilizing on-demand photon generation, optical spin read-out\cite{SpinReadOut,PurcellSpinContrast} or coalescence of indistinguishable photons\cite{NVIndistinguisability, SiVIndstinguisability}. The efficiency by which photons can be collected is, however, often compromised by the non-unity quantum yield and relatively omnidirectional emission pattern of typical QEs, whether it is a molecule, quantum dot or solid state defect\cite{SolidStateEmitters}. Fortunately, both aspects can be improved upon by engineering the photonic enviroment. Quantum yield may be increased by accelerating the radiative spontaneous decay rate, relative to intrinsic nonradiative decay, via the Purcell effect\cite{QuantumYield}. Directional emission is typically achieved by two approaches\cite{CollectionEfficiencyReview}; either a geometrical- or a mode-coupling approach. The geometrical approach relies on redirecting far-field emission by reflection or refraction on appropriately shaped surfaces, such as a parabolic mirror\cite{ParabolicMirror} or solid immersion lens\cite{SILWratchup}. Alternatively, the mode-coupling approach is based on near-field coupling QE emission to an antenna or waveguide mode. The emission pattern then conforms to that of the antenna\cite{YagiUdaScience}, while for detection with an objective, plane film waveguide modes may be redirected to free space by leakage into high index substrates\cite{PlanarDielectricAntenna} or scattering on periodic gratings\cite{DiamondMembrane, QDInMembrane}. For highly directional emission, the circular symmetric bullseye grating is particularly attractive as tight beaming of photons is achievable by appropriate grating design. Bullseye gratings have been utilized for photon collection from QEs in dielectric membranes\cite{DiamondMembrane, QDInMembrane} or situated near a grating imprinted in a metal film. Here we consider the metallic counterpart. Metallic bullseye designs currently fall in two categories. The first category consists of an aperture in a metal film, encircled by a bullseye grating\cite{BullseyeEbbesen,Brongersma,BullseyeLoncar}. This so-called bullseye aperture configuration features both large decay rate enhancement and highly directional emission from QEs situated in the aperture, however the antenna efficiency suffers from large ohmic losses. Single photon emission from such a configuration was demonstrated by Choy \textit{et. al.}, considering a nitrogen-vacancy center (NV-center) in a diamond/silver aperture\cite{BullseyeLoncar}. The design theoretically allowed for a decay rate enhancement of $\sim$ 25 (relative to bulk diamond) and collection of $\sim$ 17$\%$ of dipole emission with a 0.6 NA objective. Photon collection being compromised by reflections from the diamond/air interface appearing before the objective, and high losses of plasmon modes supported by diamond/silver interface.  More recently, a second design, consisting of a quantum dot situated in a dielectric film above a silver film bullseye grating, has been explored by Livneh \textit{et. al.}, as a way to circumvent metal losses \cite{QDInBullseye,MultipleQDInBullseye}. Exceptionally directional beaming of single photons has been demonstrated by this design, allowing for $\sim$37$\%$ of dipole emission to be collected with a 0.65 NA objective, while no decay rate enhancement is observed. The combination of appreciable decay rate enhancement and low-loss directive photon generation thus appears elusive from previous metal bullseye designs, while both are desirable to boost quantum yield and collection efficiency respectively. Further, a common characteristic of these previous bullseye designs is the corrugation of the metal film. However, recently such structuring of silver films has been observed to be problematic for single photon applications, as corrugating silver films, either by focused ion beam milling or standard electron beam lithography, result in a significant background, detrimental for single photon applications\cite{FIBbrg,QDInBullseye}. Antenna designs compatible with fabrication techniques which yield low background emission levels are therefore highly desirable for quantum photonics. Antenna designs based on planar silver films seem to be a potential approach, as background is significantly reduced compared to a corrugated surface \cite{QDInBullseye}, while planar film designs are also compatible with high-quality monocrystalline fabrication techniques\cite{SingleCrystallineSilverFilm}. While a variety of plasmonic antenna designs based on patterned metal films and/or metal nanoparticles has been realized, the planar silver film antenna employing a dielectric bullseye grating has not been explored.\\\\ 

In this study, we propose a hybrid plasmonic bullseye antenna design based on a planar silver film, employing a high-index, 100\,nm-thick, dielectric, titania (TiO$_2$) bullseye grating, situated on a low-index, 15\,nm-thin spacer layer, made of silica (SiO$_2$). The configuration combines low antenna loss and highly directional emission with a moderate decay rate enhancement. Theoretically, 85$\%$  of the QE emission may be collected by a 0.9 NA objective lens, while the photon emission rate is accelerated by a decay rate enhancement of 18 (relative to vaccum), at the design wavelength of 670\,nm. The design inherently ensures environmental protection of silver, by consequence of the protective silica layer, while the particular choice of materials allows for fabrication without compromising antenna performance by the use of lossy adhesion layers and stable operation at a laser power of 2.5\,mW corresponding to saturation of NV-center fluorescence.  For experimental demonstration of the design, a nanodiamond (ND) containing a large number of NV-centers was deterministically placed in the center of the bullseye, using an atomic force microscope. Spectrally resolving NV-center emission, a resonant enhancement peak (quality factor $\sim$ 18) is observed at the design wavelength. The on-resonance emission is selected with a band-pass filter and confirmed to be highly directional by back-focal plane imaging, with the overall emission pattern closely following numerical predictions. Specifically, emission takes the form of a radially polarized, donut-shaped beam, imposed by the rotational symmetry of the antenna, with peak emission radiated at an angle of $5^{\circ}\pm3^{\circ}$ FWHM with respect to the plane normal.
  
\section{Results and discussion}
\begin{figure}[H]
	\centering
		\includegraphics{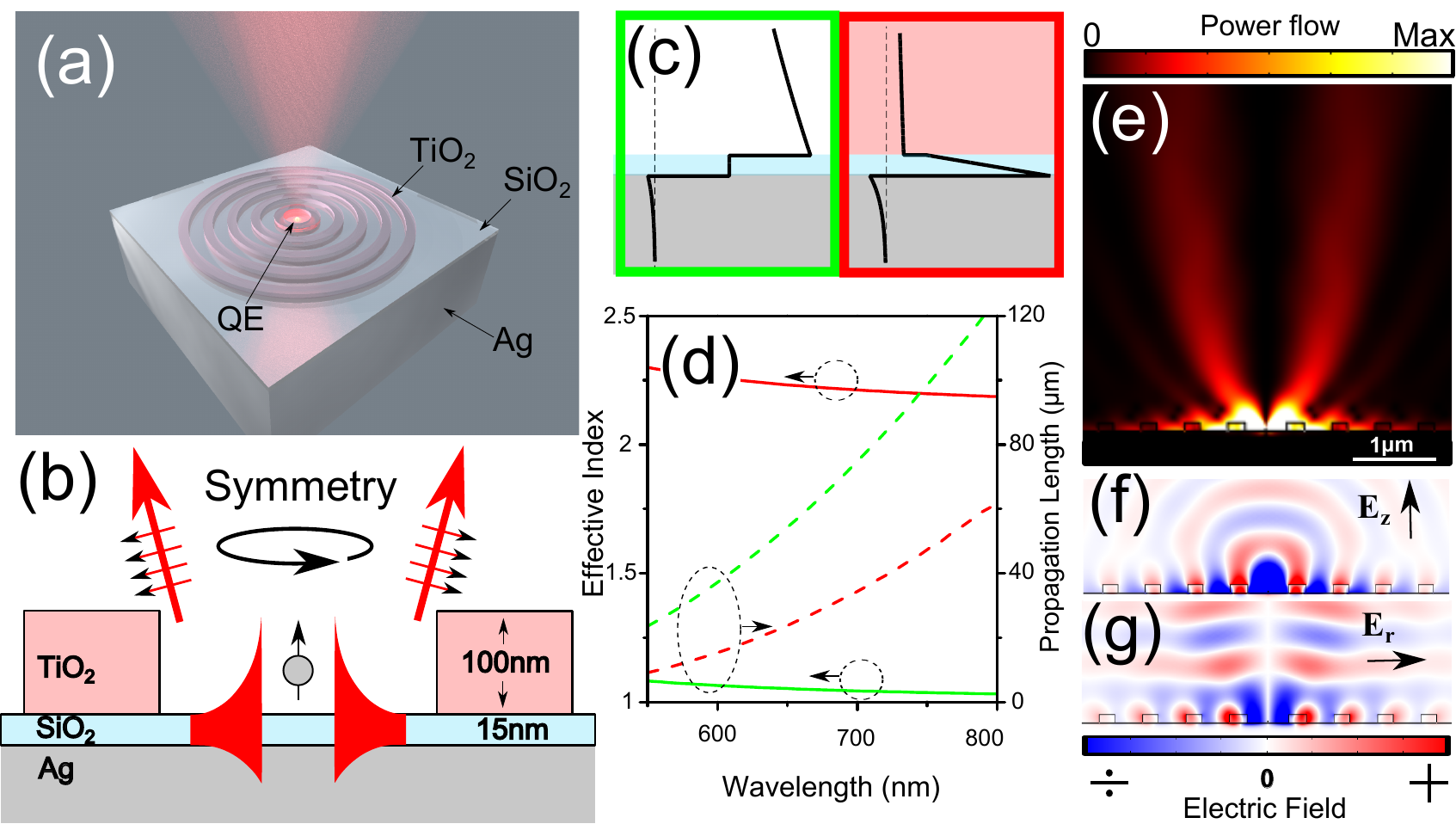}
	\caption{(a) Concept image of bullseye structure. Spontaneous emission from a centered quantum emitter is efficiently directed into the collection optics, as SPP-coupled emission scatters on the periodic ridges. (b) Profile view of the rotationally symmetric structure, emission is numerically modelled for a vertical dipole aligned with the symmetry axis. (c) Analytical out-of-plane electric field distribution for SPP modes supported by semi-infinite air (green) or TiO$_2$ (red) top layer on a 15\,nm SiO$_2$spacer, and semi-infinite bottom Ag layer. (d) Corresponding effective index (solid) and propagation length (dashed). (e) On resonance power density flow of optimized antenna with corresponding (f) out-of-plane and (g) radial electric field distributions.}
	\label{fig:Figure1}
\end{figure}

The general operation of our device is conceptually illustrated in figure \ref{fig:Figure1}(a,b). A QE, centered in the bullseye antenna in close proximity to the silver film, spontaneously decays by excitation of surface plasmon polaritons (SPP) propagating along the dielectric-metal interface. The SPP subsequently  scatter directionally on the periodically spaced TiO$_2$ ridges, resulting in highly directional emission.\\ 
The emission of a QE or electric dipole, situated in close proximity to a silver film, has been thoroughly studied\cite{Drexhage} and analytically described \cite{FordAndWeber,Chance}. It is thus well known that dipole emission is accelerated when the dipole is oriented normal to the metal plane, as emission is efficiently coupled into SPP.  On the other hand, when the dipole is oriented along the plane, emission is suppressed, as the dipole induces a mirrored anti-phase charge distribution in the silver film (see Support Information (SI)). In the present study we thus neglect the weak in-plane dipole emission, and model QE emission solely in terms of the dominating vertical dipole (figure \ref{fig:Figure1}b). Placing the dipole along the symmetry z-axis, our system reduces to a cylindrical symmetric system, for which we assume a constant azimuthal phase, as no particular phase preference can be expected. In other words, the emission field is assumed to be unchanged upon rotation about the z-axis. Clearly, in this case the radial electric field component is singular on the symmetry axis, and must therefore be zero (figure \ref{fig:Figure1}g). The condition prohibits emission along the symmetry axis, as the transverse field of a plane wave, propagating along the symmetry axis, must be zero. The rotational symmetry of dipole source and antenna is thus expected to result in a radially polarized emission field, for which emission normal to the silver film is prohibited, regardless of emission wavelength.\\
With the symmetry constraints in place, we now seek to maximize dipole emission into the 0.9 NA objective, at the target wavelength of $\lambda_0=$670\,nm, by introducing a dielectric grating for scattering of SPP coupled emission. In order to maximize antenna efficiency, we seek to minimize SPP propagation loss by employing a high-index TiO$_2$ grating (n=2.2 index, measured with ellipsometry see SI), separated from the silver film by a thin low-index  SiO$_2$ spacer layer (n=1.45 index). The SPP mode of the high index-low index spacer - on conductor configuration\cite{High_Low_Index_SPPTheory} is recieving increasing interest as a possible approach for low-loss plasmonics\cite{EvelynHu_NVCavity,LowLossPlasmonics}. Analytical solutions for the SPP mode supported by the 3-layer structure (figure \ref{fig:Figure1} c,d) reveal large modulation of the effective index from $N_{eff}^{Air}=$1.05 (Air-SiO$_2$-Ag) to $N_{eff}^{TiO2}=$2.25 (TiO$_2$-SiO$_2$-Ag), with propagation lengths of respectively L$^{Air}_p$=63\,$\mu{m}$ and L$^{TiO2}_p$=28\,$\mu{m}$ far exceeding lateral antenna dimensions. SPP coupled emission is thus prone to either scatter on the TiO$_2$ grating or reflect back to the emitter, while propagation losses are minimal. For fabrication purposes, we set titania and silica film thicknesses to 100\,nm and 15\,nm respectively, and numerically optimize in-plane grating parameters for collecting the maximal amount of dipole power with a 0.9 NA objective (see SI for optimization procedure). The inner TiO$_2$ ridge (inner radius 250\,nm, TiO$_2$ width 212\,nm) in this case forms a standing wave cavity (figure \ref{fig:Figure1}f), accelerating dipole emission by a decay rate enhancement of 18 (relative to vacuum). Simultaneously, the periodic TiO$_2$ grating (period 520\,nm, duty cycle 0.36), directionally scatters SPP coupled emission, such that 85$\%$ of the power emitted by the dipole is collected by the objective. While on-resonance SPP emission is scattered at an angle near normal to silver plane (figure \ref{fig:Figure1} e,g), the scattering angle generally seems to follow the grating equation (see SI for comparison). The geometrical optimization of TiO$_2$ height is a trade-off between decay rate enhancement and collection efficiency. Collection efficiency defined as the fraction of the total dipole power, collected by the objective, while decay rate enhancement is the factor by which the total dipole emission power increases in the bullseye environment, relative to vaccum.
 Indeed, a significant increase in decay rate enhancement is possible by increasing the TiO$_2$ thickness (see SI for modelling) at the cost of collection efficiency, as the SPP perform an increasing number of lossy round trips in the cavity before scattering to free space. For  practical fabrication purposes, the low aspect ratio design of  100\,nm TiO$_2$ was preferred, while the 15\,nm SiO$_2$ film thickness was set to ensure a homogenous coverage of the silver film, for environmental protection of the silver.\\\\

\begin{figure}[H]
	\centering
		\includegraphics{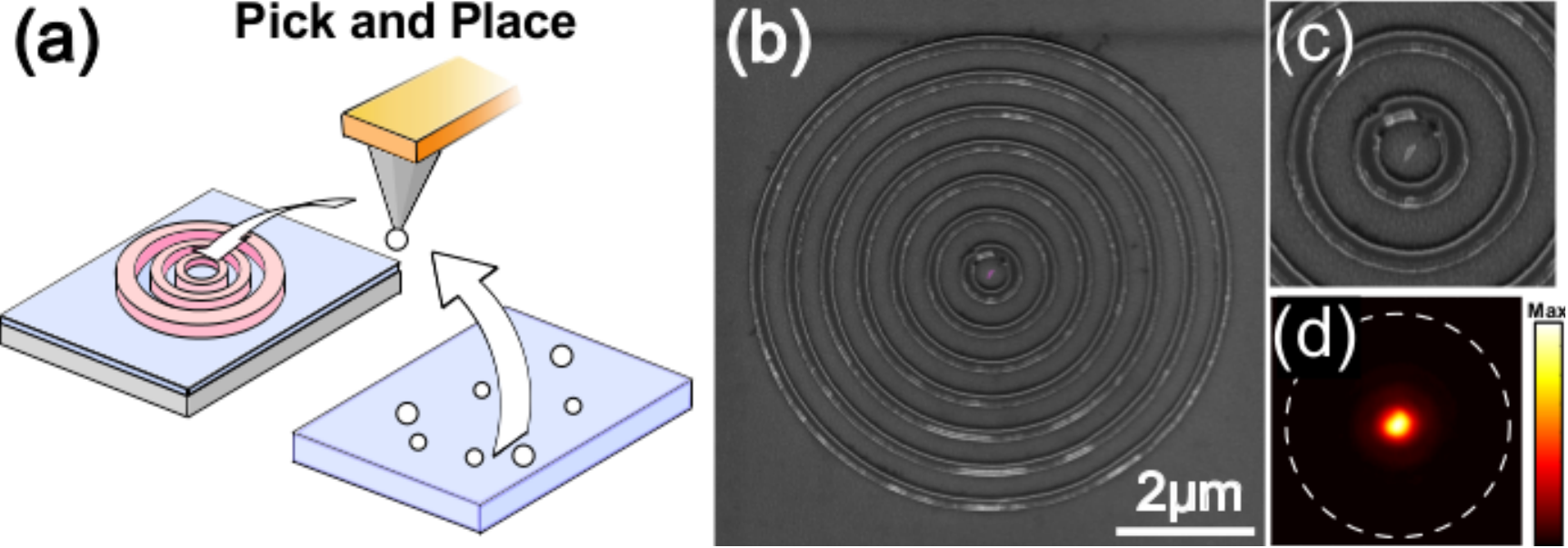}
		\caption{(a) Sketch of AFM "pick and place" technique used in transferring a ND from a coverslip to the bullseye center. (b) Electron micrograph of fabricated bullseye antenna, the transferred ND appears in false color in the center, (c) clearly apparent in the zoomed image. (d) False color CCD image of emission from the ND, dashed line indicates bullseye boundary. }
	\label{fig:Figure2}
\end{figure}

\begin{figure}[H]
	\centering
		\includegraphics{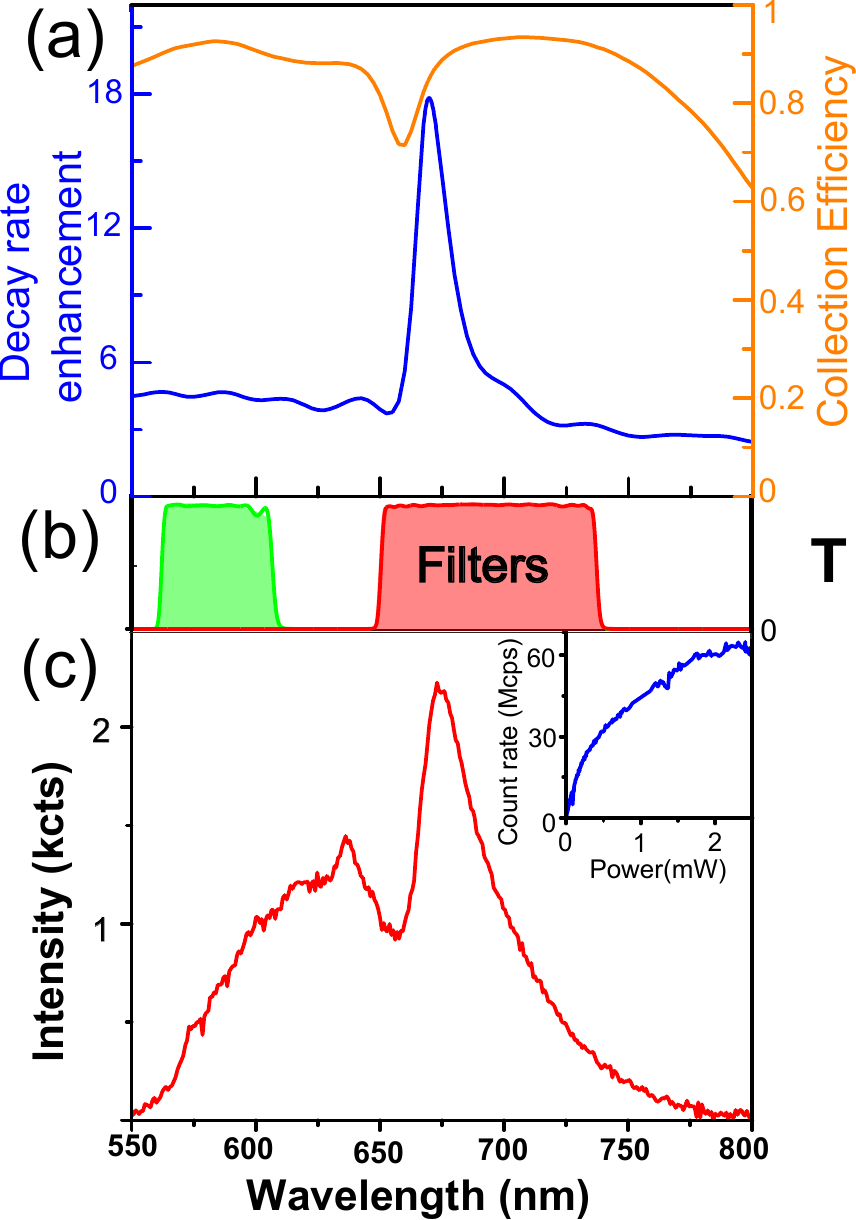}
		\caption{(a) Modelled decay rate enhancement (blue) and collection efficiency (orange) for fabricated bullseye design. (b) Transimission of bandpass filters used for detecting off-resonance (560-610\,nm) (green) and on-resonance (650-740\,nm) (red) emission, respectively. (c) Emission spectrum from ND centered in bullseye. Inset shows the background corrected saturation curve, demonstrating stable antenna operation up to saturation laser power.}
	\label{fig:Figure3}
\end{figure}

\begin{figure}[H]
	\centering
		\includegraphics{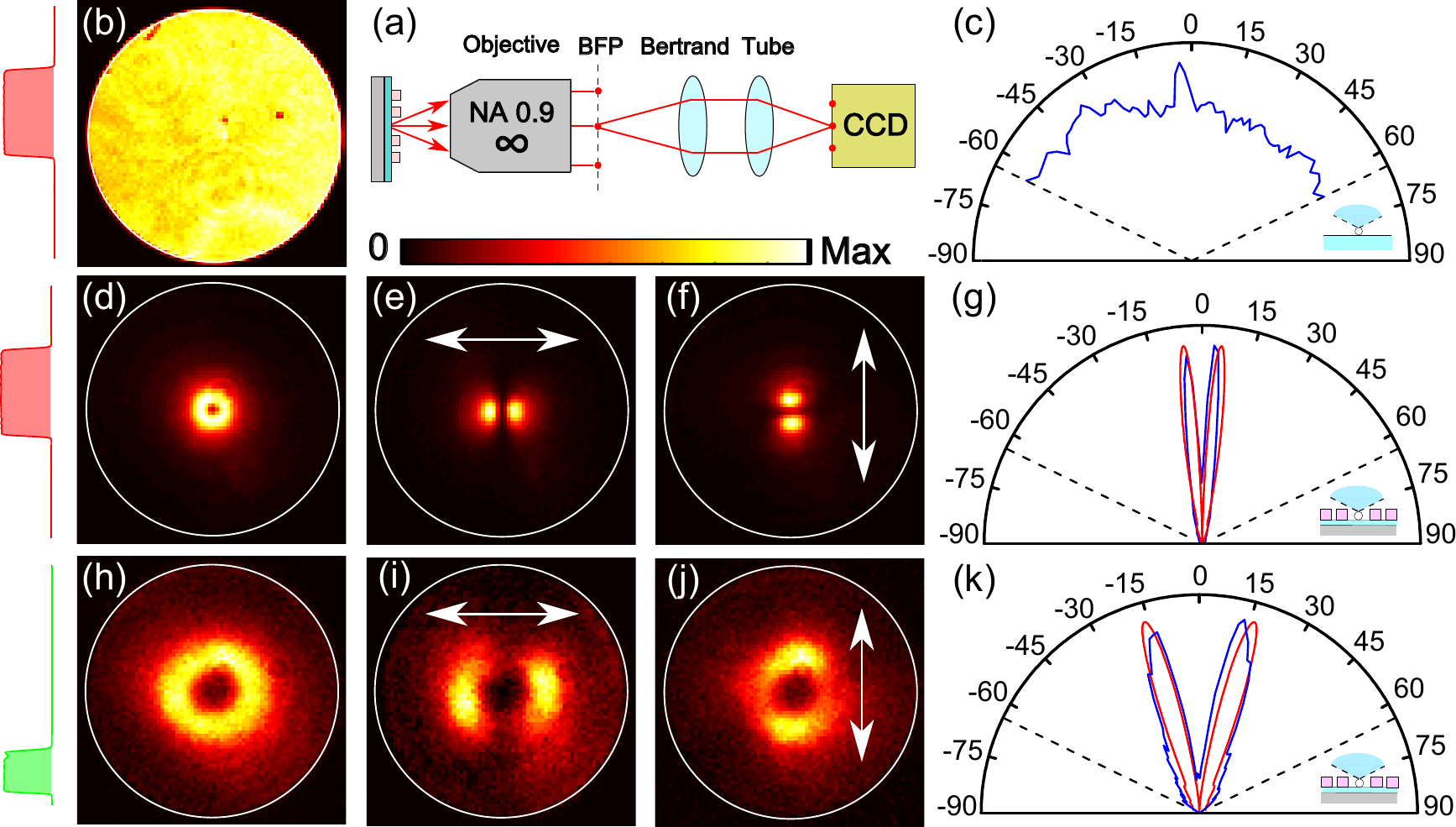}
		\caption{(a) Schematic of experimental setup for back-focal plane (BFP) imaging. (b) Reference BFP-image from ND on coverslip and (c) corresponding emission pattern. The detection limit of the 0.9 numerical aperture of the objective is indicated by respectively a white circle or black dashed lines. BFP-images filtered for (d-f) on- and (h-j) off-resonance emission from ND in bullseye, Imaged with (e, i) horizontal or (f, j) vertical analyzer. Experimental (blue) and modelled (red) emission patterns for (g) on-resonance , model wavelength 670\,nm and (k) off-resonance regime, model wavelength 610\,nm.}
	\label{fig:Figure4}
\end{figure}

The bullseye antenna sample was fabricated on a Si substrate by e-beam evaporation of 10\,nm Ti, 3\,nm Ge followed by a 200\,nm Ag film and topped by a 15\,nm SiO$_2$ layer, without breaking the vacuum. The 100\,nm TiO$_2$ bullseye grating was subsequently formed by standard electron beam lithography and e-beam evaporation (figure \ref{fig:Figure2}b). Importantly, device performance was not degraded by the use of any lossy adhesion layers at the Ag-SiO$_2$-TiO$_2$ interfaces, while remaining mechanically robust to sonication, applied during lift-off.  A ND (length $\sim$ 140\,nm , height $\sim$ 35\,nm) containing a large number of NV-centers was subsequently picked up from a coverslip and placed in the center of the bullseye antenna, using an atomic force microscope (AFM) (figure \ref{fig:Figure2}a). The AFM "pick and place" technique\cite{PickAndPlace} allows for precise centering of the ND, as confirmed by electron microscopy (figure \ref{fig:Figure2}d). Pumping the ND with a 532\,nm continuous-wave laser, we image the NV-center fluorescence onto a CCD camera, using a 0.9 NA x100 objective and fluorescence filtering from a dichroic mirror (cut-off 550\,nm) and long pass filter (cut-off 550\,nm) (see SI for schematic of setup). The NV-center emission is observed as a spot, tightly confined to the center of bullseye antenna, indicating efficient scattering of SPP-coupled emission on the TiO$_2$ ridges (figure \ref{fig:Figure2}c). Correspondingly, antennas without a ND appear dark, when mapping fluorescence by laser-scanning confocal microscopy (see SI for comparison). In order to observe the wavelength dependent decay rate enhancement, expected from numerical modelling (figure \ref{fig:Figure3}a), the NV-center emission was spectrally resolved on a grating spectrometer (Figure \ref{fig:Figure3}c). The zero phonon lines (ZPL) for the neutral- (575\,nm) and negative (637\,nm) charge states confirm the NV-center as the origin of fluorescence, while the overall spectrum is dominated by a resonant enhancement peak at 675\,nm, in reasonable agreement with the modelled decay rate enhancement. The experimental quality factor of $\sim$ 18 (estimated from spectrum) is broader than the modelled resonance $\sim$ 41 (estimated from decay rate enhancement curve), presumably due to fabrication imperfections. Collection efficiency is not expected to significantly affect the spectrum, as modelling finds overall high broadband collections efficiency (figure \ref{fig:Figure3}a), stemming from emission into SPP and grating scattering of SPP into objective both being broadband effects. \\
For efficient photon generation, it is typically desired to maximize photon rate from the QE, by pumping the QE to saturation. We confirmed that our device is stable under such conditions by increasing laser power to the 2.5\,mW limit of our equipment. Simultaneously monitoring the photon rate detected by a avalanche photo diode, we smoothly transition to stable operation at saturation (figure \ref{fig:Figure3}c inset). Subsequently reducing laser power, we retrace the saturation curve, thereby confirming no antenna alterations. The planar film bullseye antenna is generally expected to exhibit a higher laser damage threshold, than structured film configurations, as nanostructuring of metals suffer from melting point depression\cite{MeltingPointDepression}. Having confirmed stable operation at saturation laser power and the presence of resonantly enhanced emission, we proceed by examining the off- and on-resonance emission pattern, selected with bandpass filter in the wavelength range 560-610\,nm and 650-740\,nm (figure \ref{fig:Figure3}b). Introducing a bertrand lens in infinity space, we resolve the emission pattern by imaging the back-focal plane(BFP) of the objective onto the CCD camera (figure \ref{fig:Figure4}a). This lens configuration is particularly well suited for fourier microscopy\cite{FourierMicroscopy}. Filtering for resonant emission reveals a highly directional radiation pattern in the characteristic donut-shaped pattern, imposed by the antenna symmetry (figure \ref{fig:Figure4}d). Introducing an analyzer in the optical path, the radiation is confirmed to be radially polarized (figure \ref{fig:Figure4} e,f). This is apparent as emission is spatially extinquished along the axis normal to the analyzer axis for a random analyzer orientation.  For a comparison with the numerical design, we extracted the angular radiation pattern from a single pixel slice of the BFP-image, finding the experiment generally replicates numerical expectations. On resonance, the peak emission is detected at an angle of 5$^{\circ} \pm3^{\circ}$ FWHM wrt. the plane normal, in good agreement with the designed emission angle of 6$^{\circ} \pm4^{\circ}$ FWHM, modelled for the resonance wavelength of 670\,nm (figure \ref{fig:Figure4}g). Probing off-resonant emission, we find the peak emission increasing to an angle of 14$^{\circ} \pm8^{\circ}$ FWHM (figure \ref{fig:Figure4}k), while the symmetry conditioned, radially polarized donut shaped beam pattern is conserved (figure \ref{fig:Figure4}h-j). Off-resonance emission is in line with the modelled emission for a wavelength of 610\,nm having peak emission at an angle of 16$^{\circ} \pm5^{\circ}$. Off-resonance emission is modelled at the wavelength of 610\,nm, as experimentally, the maximum spectral power transmitted by the band-pass filter is found at this wavelength (figure \ref{fig:Figure3}c). 
The directional emission is in stark contrast to the omni-directional emission of the same type of ND, when situated on a coverslip (figure \ref{fig:Figure4}b,c).    
\\\\
The relatively narrowband decay rate enhancement and directive emission, demonstrated for this design, may be of interest for indistinguisable photon experiments, as the selective enhancement and efficient collection of ZPL photons from solid state emitters is desirable. The simple fabrication allows for easy tuning of the antenna resonance to the ZPL of new promising QEs such as the silicon- or germanium vacancy centers \cite{SiV,GeV}. However, for scalable single photon source fabrication, large-scale QE positioning techniques need to be explored. Potential approaches may be lithographic patterning\cite{HolePLacement} or electro-static pad positioning\cite{ElectrostaticAssembly}. Further, the highly directional emission pattern should lend itself to optical fiber coupling. Theoretically, 23$\%$ and 74$\%$ of dipole emission (wavelength 670\,nm) fall within the numerical aperture of respectively a single- (NA 0.12) or multi-mode fiber (NA 0.4).  It is in this context worth noting that the radially polarized emission pattern, inherent to the bullseye antenna, may be converted to the fundamental mode of an optical fiber with high fidelity\cite{FiberModeConv}. Directive emission is further advantageous for NV ensemble-based sensing applications, relying on optical read-out of the NV-center spin state, where strong decay rate enhancement should be avoided.\cite{PurcellSpinContrast}  The design may potentially be optimized for operation in an oil index matching environment, to limit reflection losses at the fiber or objective input.

\section{Conclusion}
In summary, we have proposed a high-efficiency hybrid plasmonic bullseye antenna design, consisting of a high-index titania grating, separated from a planar silver film by a thin low-index silica spacer layer. The architecture is motivated by previous issues with background emission resulting from nanostructuring of silver films. Our design combines low antenna loss, directional emission and moderate decay rate enhancement, leading to a theoretical collection efficiency of 85\% for a 0.9 NA objective and acceleration of the emission rate by a decay rate enhancement of 18, at the design wavelength of 670\,nm. The design is experimentally realized by standard electron beam lithography, followed by deterministic placement of a fluorescent ND in the bullseye, using the AFM "pick and place" technique. The particular material design allows for fabrication without lossy adhesion layers compromising device performance, and stable operation of the antenna at laser powers large enough for saturated pumping of the NV-centers contained in the ND. Decay rate enhancement is experimentally observed as a resonant peak in the NV-center emission spectrum at 675\,nm (quality factor $\sim$18).  The resonantly filtered fluorescence is demonstrated to be highly directional by back-focal plane imaging, with peak emission at an angle of  5$^{\circ}$ with respect to plane normal. Specifically, the emission pattern takes the form of a donut-shaped, radially polarized beam as imposed by the antenna symmetry, regardless of emission wavelength. Experimental observations closely mimic the numerical design, thereby validating the proposed design. The demonstrated design is significant for quantum technological development, as light-matter interfaces ensuring efficient photon collection from single QEs are highly desirable for quantum optical applications.

\section*{Methods}
\subsection*{Sample Preparation}
The bullseye antenna was fabricated by successive electron-beam evaporation of 10\,nm Ti, 3\,nm Ge, 200\,nm Ag topped by 15\,nm SiO$_2$ on a Si wafer, at a deposition rate of $\sim$1 \AA /s and $10^{-6}$\,mbar chamber pressure, without breaking the vacuum.  PMMA 4A was spincoated at 4000\,rpm on the sample, and prebaked at 180\,C$^o$ for 3\,min before patterning by a 100kV electron beam lithography system (Leica VB6). After development (1:3 MIBK-to-IPA for 1\,min followed by 1\,min rinse in IPA), a 100\,nm TiO$_2$ layer was deposited by electron-beam evaporation at  $\sim$1 \AA /s and $10^{-6}$\,mbar pressure. Lift-off in acetone at 60\,$C^{o}$ for 5\,h was partially successful, as subsequently 5\,min of sonication in acetone was required to remove PMMA between the TiO$_2$ ridges to reveal the bullseye. In preparation of ND placement, 100\,nm ND's containing $\sim$ 400 NV-centers (Adamas technology) was spincoated on a coverslip, previously cleaned in piraniha etch (Nanostrip x2 - KMG electronic chemicals). To ease the transfer, the bullseye sample was coated with a $\sim$ 2\,nm thick layer of positively charged poly-allylamine hydrochloride (PAH) layer. ND pick-up was performed by a force curve sequence, with the AFM cantilever situated above the ND. Successful pick-up was confirmed by a subsequent non-contact scan. Exchanging samples, a force curve was performed in the center of the bullseye, followed by an AFM scan to confirm the placement of the ND.

\subsection*{Numerical Modelling}
Numerical modelling was performed in the commercially available Comsol Multiphysics 5.1. The full 3-D field of the axial symmetric system is numerically modelled by solving a 2-D slice, on the assumption of a constant azimuthal phase. The    limited computational requirements of 2-D modelling, allowed us to model the full bullseye antenna, in a domain of radius 22$\mu{m}$, bounded from the top by a perfectly matched layer and below by the silver film. Material parameters for silver were obtained for tabulated data \cite{JohnsonChristy}, while TiO$_2$ data was measured by ellipsometry (see SI) and a refractive index of 1.45 was set for SiO$_2$. The emission field was generated by an electric dipole source positioned on the symmetry axis, 15\,nm above the SiO$_2$ film. The decay rate enhancement was obtained as the total power dissipated by the dipole in the bullseye environment, relative to that of free space. Collection efficiency was obtained as the power integrated over a 0.9 NA collection surface, relative to the total power dissipated by the dipole. The radiation pattern was obtained by plotting the Poynting vector over a spherical arc in the 2-D plane.

\subsection*{Experiment}
The ND was excited with a 532\,nm continous wave laser, focused onto the sample by a 0.9 NA x100 objective which was mounted on a piezo-stage for fine positioning of laserspot. Fluorescence collected by the same objective was filtered by a 550\,nm dichroic mirror (DMLP550L-Thorlabs) and 550\,nm long pass filter (FEL0550- Thorlabs), before being directed to a spectrometer (QE65000 - Ocean Optics) or avalance photodiode (SPCM-AQRH - Excelitas) in a confocal detection configuration using a 50\,$\mu{m}$ pinhole. Alternatively fluorescence was detected by a charge coupled camera (414Ex - Atik Cameras) imaging sample plane (using a 20\,mm tube lens) or back-focal plane (600\,mm bertrand- and 75\,mm tube lens) onto the camera. off- or on-resonance emission was selected by filters of respectively 560-610\,nm and 650-740\,nm transmission bands, while an analyzer (LPVISC050-MP2 - Thorlabs) was introduced to probe polarization. Back-focal plane images from bullseye antenna, was background corrected for a corresponding image from an empty antenna. Saturation curve measurement was performed with a ND2 filter positioned infront of the avalance photodiode, the presented count rate was corrected for transmission of ND2 filter and the pinhole. See SI for a schematic of experimental setup.  
\begin{acknowledgement}
The authours gratefully acknowledge the finanical support of the European Research Council (Grant 341054 (PLAQNAP)), Georgia St. (Grant 106806), MRSEC (Grant 105675), ONR-DURIP (Grant No. N00014-16-1-2767), 
AFOSR-MURI (Grant No. FA9550-14-10389) and DOE (Grant DE-SC0017717).

\end{acknowledgement}
 
\begin{suppinfo}
The following files are available free of charge.
S1: Analytical model of electric dipole above silver film. S2: Ellipsometry measurement of TiO$_2$ refractive index. S3: Numerical optimization procedure of bullseye antenna. S4: Comparision of bullseye emission angle with grating equation. S5: Optimized antenna design for different TiO$_2$ heights. S6: Schematic of experimental setup. S7: Confocal scan of bullseye antenna with- and without ND. 
\end{suppinfo}


\providecommand{\latin}[1]{#1}
\makeatletter
\providecommand{\doi}
  {\begingroup\let\do\@makeother\dospecials
  \catcode`\{=1 \catcode`\}=2\doi@aux}
\providecommand{\doi@aux}[1]{\endgroup\texttt{#1}}
\makeatother
\providecommand*\mcitethebibliography{\thebibliography}
\csname @ifundefined\endcsname{endmcitethebibliography}
  {\let\endmcitethebibliography\endthebibliography}{}


\begin{mcitethebibliography}{37}
\providecommand*\natexlab[1]{#1}
\providecommand*\mciteSetBstSublistMode[1]{}
\providecommand*\mciteSetBstMaxWidthForm[2]{}
\providecommand*\mciteBstWouldAddEndPuncttrue
  {\def\EndOfBibitem{\unskip.}}
\providecommand*\mciteBstWouldAddEndPunctfalse
  {\let\EndOfBibitem\relax}
\providecommand*\mciteSetBstMidEndSepPunct[3]{}
\providecommand*\mciteSetBstSublistLabelBeginEnd[3]{}
\providecommand*\EndOfBibitem{}
\mciteSetBstSublistMode{f}
\mciteSetBstMaxWidthForm{subitem}{(\alph{mcitesubitemcount})}
\mciteSetBstSublistLabelBeginEnd
  {\mcitemaxwidthsubitemform\space}
  {\relax}
  {\relax}

\bibitem[O'Brien \latin{et~al.}(2009)O'Brien, Furusawa, and
  Vuckovic]{QuantumTechnology}
O'Brien,~J.~L.; Furusawa,~A.; Vuckovic,~J. Photonic quantum technologies.
  \emph{Nat Photon} \textbf{2009}, \emph{3}, 687--695\relax
\mciteBstWouldAddEndPuncttrue
\mciteSetBstMidEndSepPunct{\mcitedefaultmidpunct}
{\mcitedefaultendpunct}{\mcitedefaultseppunct}\relax
\EndOfBibitem
\bibitem[Steiner \latin{et~al.}(2010)Steiner, Neumann, Beck, Jelezko, and
  Wrachtrup]{SpinReadOut}
Steiner,~M.; Neumann,~P.; Beck,~J.; Jelezko,~F.; Wrachtrup,~J. Universal
  enhancement of the optical readout fidelity of single electron spins at
  nitrogen-vacancy centers in diamond. \emph{Phys. Rev. B} \textbf{2010},
  \emph{81}, 035205\relax
\mciteBstWouldAddEndPuncttrue
\mciteSetBstMidEndSepPunct{\mcitedefaultmidpunct}
{\mcitedefaultendpunct}{\mcitedefaultseppunct}\relax
\EndOfBibitem
\bibitem[Bogdanov \latin{et~al.}(2017)Bogdanov, Shalaginov, Akimov, Lagutchev,
  Kapitanova, Liu, Woods, Ferrera, Belov, Irudayaraj, Boltasseva, and
  Shalaev]{PurcellSpinContrast}
Bogdanov,~S.; Shalaginov,~M.~Y.; Akimov,~A.; Lagutchev,~A.~S.; Kapitanova,~P.;
  Liu,~J.; Woods,~D.; Ferrera,~M.; Belov,~P.; Irudayaraj,~J.; Boltasseva,~A.;
  Shalaev,~V.~M. Electron spin contrast of Purcell-enhanced nitrogen-vacancy
  ensembles in nanodiamonds. \emph{Phys. Rev. B} \textbf{2017}, \emph{96},
  035146\relax
\mciteBstWouldAddEndPuncttrue
\mciteSetBstMidEndSepPunct{\mcitedefaultmidpunct}
{\mcitedefaultendpunct}{\mcitedefaultseppunct}\relax
\EndOfBibitem
\bibitem[Bernien \latin{et~al.}(2012)Bernien, Childress, Robledo, Markham,
  Twitchen, and Hanson]{NVIndistinguisability}
Bernien,~H.; Childress,~L.; Robledo,~L.; Markham,~M.; Twitchen,~D.; Hanson,~R.
  Two-Photon Quantum Interference from Separate Nitrogen Vacancy Centers in
  Diamond. \emph{Phys. Rev. Lett.} \textbf{2012}, \emph{108}, 043604\relax
\mciteBstWouldAddEndPuncttrue
\mciteSetBstMidEndSepPunct{\mcitedefaultmidpunct}
{\mcitedefaultendpunct}{\mcitedefaultseppunct}\relax
\EndOfBibitem
\bibitem[Sipahigil \latin{et~al.}(2014)Sipahigil, Jahnke, Rogers, Teraji,
  Isoya, Zibrov, Jelezko, and Lukin]{SiVIndstinguisability}
Sipahigil,~A.; Jahnke,~K.~D.; Rogers,~L.~J.; Teraji,~T.; Isoya,~J.;
  Zibrov,~A.~S.; Jelezko,~F.; Lukin,~M.~D. Indistinguishable Photons from
  Separated Silicon-Vacancy Centers in Diamond. \emph{Phys. Rev. Lett.}
  \textbf{2014}, \emph{113}, 113602\relax
\mciteBstWouldAddEndPuncttrue
\mciteSetBstMidEndSepPunct{\mcitedefaultmidpunct}
{\mcitedefaultendpunct}{\mcitedefaultseppunct}\relax
\EndOfBibitem
\bibitem[Aharonovich \latin{et~al.}(2016)Aharonovich, Englund, and
  Toth]{SolidStateEmitters}
Aharonovich,~I.; Englund,~D.; Toth,~M. Solid-state single-photon emitters.
  \emph{Nat Photon} \textbf{2016}, \emph{10}, 631--641\relax
\mciteBstWouldAddEndPuncttrue
\mciteSetBstMidEndSepPunct{\mcitedefaultmidpunct}
{\mcitedefaultendpunct}{\mcitedefaultseppunct}\relax
\EndOfBibitem
\bibitem[Kinkhabwala \latin{et~al.}(2009)Kinkhabwala, Yu, Fan, Avlasevich,
  Mullen, and E.]{QuantumYield}
Kinkhabwala,~A.; Yu,~Z.; Fan,~S.; Avlasevich,~Y.; Mullen,~K.; E.,~M. Large
  single-molecule fluorescence enhancements produced by a bowtie nanoantenna.
  \emph{Nat Photon} \textbf{2009}, \emph{3}, 654--657\relax
\mciteBstWouldAddEndPuncttrue
\mciteSetBstMidEndSepPunct{\mcitedefaultmidpunct}
{\mcitedefaultendpunct}{\mcitedefaultseppunct}\relax
\EndOfBibitem
\bibitem[Barnes \latin{et~al.}(2002)Barnes, Bj{\"o}rk, G{\'e}rard, Jonsson,
  Wasey, Worthing, and Zwiller]{CollectionEfficiencyReview}
Barnes,~W.; Bj{\"o}rk,~G.; G{\'e}rard,~J.; Jonsson,~P.; Wasey,~J.;
  Worthing,~P.; Zwiller,~V. Solid-state single photon sources: light collection
  strategies. \emph{The European Physical Journal D - Atomic, Molecular,
  Optical and Plasma Physics} \textbf{2002}, \emph{18}, 197--210\relax
\mciteBstWouldAddEndPuncttrue
\mciteSetBstMidEndSepPunct{\mcitedefaultmidpunct}
{\mcitedefaultendpunct}{\mcitedefaultseppunct}\relax
\EndOfBibitem
\bibitem[Schell \latin{et~al.}(2014)Schell, Neumer, Shi, Kaschke, Fischer,
  Wegener, and Benson]{ParabolicMirror}
Schell,~A.~W.; Neumer,~T.; Shi,~Q.; Kaschke,~J.; Fischer,~J.; Wegener,~M.;
  Benson,~O. Laser-written parabolic micro-antennas for efficient photon
  collection. \emph{Applied Physics Letters} \textbf{2014}, \emph{105},
  231117\relax
\mciteBstWouldAddEndPuncttrue
\mciteSetBstMidEndSepPunct{\mcitedefaultmidpunct}
{\mcitedefaultendpunct}{\mcitedefaultseppunct}\relax
\EndOfBibitem
\bibitem[Jamali \latin{et~al.}(2014)Jamali, Gerhardt, Rezai, Frenner, Fedder,
  and Wrachtrup]{SILWratchup}
Jamali,~M.; Gerhardt,~I.; Rezai,~M.; Frenner,~K.; Fedder,~H.; Wrachtrup,~J.
  Microscopic diamond solid-immersion-lenses fabricated around single defect
  centers by focused ion beam milling. \emph{Review of Scientific Instruments}
  \textbf{2014}, \emph{85}, 123703\relax
\mciteBstWouldAddEndPuncttrue
\mciteSetBstMidEndSepPunct{\mcitedefaultmidpunct}
{\mcitedefaultendpunct}{\mcitedefaultseppunct}\relax
\EndOfBibitem
\bibitem[Curto \latin{et~al.}(2010)Curto, Volpe, Taminiau, Kreuzer, Quidant,
  and van Hulst]{YagiUdaScience}
Curto,~A.~G.; Volpe,~G.; Taminiau,~T.~H.; Kreuzer,~M.~P.; Quidant,~R.; van
  Hulst,~N.~F. Unidirectional Emission of a Quantum Dot Coupled to a
  Nanoantenna. \emph{Science} \textbf{2010}, \emph{329}, 930--933\relax
\mciteBstWouldAddEndPuncttrue
\mciteSetBstMidEndSepPunct{\mcitedefaultmidpunct}
{\mcitedefaultendpunct}{\mcitedefaultseppunct}\relax
\EndOfBibitem
\bibitem[Lee \latin{et~al.}(2011)Lee, Chen, Eghlidi, Kukura, Lettow, Renn,
  Sandoghdar, and Gotzinger]{PlanarDielectricAntenna}
Lee,~K.~G.; Chen,~X.~W.; Eghlidi,~H.; Kukura,~P.; Lettow,~R.; Renn,~A.;
  Sandoghdar,~V.; Gotzinger,~S. A planar dielectric antenna for directional
  single-photon emission and near-unity collection efficiency. \emph{Nat
  Photon} \textbf{2011}, \emph{5}, 166--169\relax
\mciteBstWouldAddEndPuncttrue
\mciteSetBstMidEndSepPunct{\mcitedefaultmidpunct}
{\mcitedefaultendpunct}{\mcitedefaultseppunct}\relax
\EndOfBibitem
\bibitem[Li \latin{et~al.}(2015)Li, Chen, Zheng, Mouradian, Dolde,
  Schr{\"o}der, Karaveli, Markham, Twitchen, and Englund]{DiamondMembrane}
Li,~L.; Chen,~E.~H.; Zheng,~J.; Mouradian,~S.~L.; Dolde,~F.; Schr{\"o}der,~T.;
  Karaveli,~S.; Markham,~M.~L.; Twitchen,~D.~J.; Englund,~D. Efficient Photon
  Collection from a Nitrogen Vacancy Center in a Circular Bullseye Grating.
  \emph{Nano Letters} \textbf{2015}, \emph{15}, 1493--1497\relax
\mciteBstWouldAddEndPuncttrue
\mciteSetBstMidEndSepPunct{\mcitedefaultmidpunct}
{\mcitedefaultendpunct}{\mcitedefaultseppunct}\relax
\EndOfBibitem
\bibitem[Davan{\c c}o \latin{et~al.}(2011)Davan{\c c}o, Rakher, Schuh,
  Badolato, and Srinivasan]{QDInMembrane}
Davan{\c c}o,~M.; Rakher,~M.~T.; Schuh,~D.; Badolato,~A.; Srinivasan,~K. A
  circular dielectric grating for vertical extraction of single quantum dot
  emission. \emph{Applied Physics Letters} \textbf{2011}, \emph{99},
  041102\relax
\mciteBstWouldAddEndPuncttrue
\mciteSetBstMidEndSepPunct{\mcitedefaultmidpunct}
{\mcitedefaultendpunct}{\mcitedefaultseppunct}\relax
\EndOfBibitem
\bibitem[Aouani \latin{et~al.}(2011)Aouani, Mahboub, Bonod, Devaux, Popov,
  Rigneault, Ebbesen, and Wenger]{BullseyeEbbesen}
Aouani,~H.; Mahboub,~O.; Bonod,~N.; Devaux,~E.; Popov,~E.; Rigneault,~H.;
  Ebbesen,~T.~W.; Wenger,~J. Bright Unidirectional Fluorescence Emission of
  Molecules in a Nanoaperture with Plasmonic Corrugations. \emph{Nano Letters}
  \textbf{2011}, \emph{11}, 637--644\relax
\mciteBstWouldAddEndPuncttrue
\mciteSetBstMidEndSepPunct{\mcitedefaultmidpunct}
{\mcitedefaultendpunct}{\mcitedefaultseppunct}\relax
\EndOfBibitem
\bibitem[Jun \latin{et~al.}(2011)Jun, Huang, and Brongersma]{Brongersma}
Jun,~Y.~C.; Huang,~K.~C.; Brongersma,~M.~L. Plasmonic beaming and active
  control over fluorescent emission. \emph{ncomms} \textbf{2011}, \emph{2},
  1--6\relax
\mciteBstWouldAddEndPuncttrue
\mciteSetBstMidEndSepPunct{\mcitedefaultmidpunct}
{\mcitedefaultendpunct}{\mcitedefaultseppunct}\relax
\EndOfBibitem
\bibitem[Choy \latin{et~al.}(2013)Choy, Bulu, Hausmann, Janitz, Huang, and
  Loncar]{BullseyeLoncar}
Choy,~J.~T.; Bulu,~I.; Hausmann,~B. J.~M.; Janitz,~E.; Huang,~I.-C.; Loncar,~M.
  Spontaneous emission and collection efficiency enhancement of single emitters
  in diamond via plasmonic cavities and gratings. \emph{Applied Physics
  Letters} \textbf{2013}, \emph{103}, 161101\relax
\mciteBstWouldAddEndPuncttrue
\mciteSetBstMidEndSepPunct{\mcitedefaultmidpunct}
{\mcitedefaultendpunct}{\mcitedefaultseppunct}\relax
\EndOfBibitem
\bibitem[Livneh \latin{et~al.}(2016)Livneh, Harats, Istrati, Eisenberg, and
  Rapaport]{QDInBullseye}
Livneh,~N.; Harats,~M.~G.; Istrati,~D.; Eisenberg,~H.~S.; Rapaport,~R. Highly
  Directional Room-Temperature Single Photon Device. \emph{Nano Letters}
  \textbf{2016}, \emph{16}, 2527--2532\relax
\mciteBstWouldAddEndPuncttrue
\mciteSetBstMidEndSepPunct{\mcitedefaultmidpunct}
{\mcitedefaultendpunct}{\mcitedefaultseppunct}\relax
\EndOfBibitem
\bibitem[Livneh \latin{et~al.}(2015)Livneh, Harats, Yochelis, Paltiel, and
  Rapaport]{MultipleQDInBullseye}
Livneh,~N.; Harats,~M.~G.; Yochelis,~S.; Paltiel,~Y.; Rapaport,~R. Efficient
  Collection of Light from Colloidal Quantum Dots with a Hybrid Metal
  Dielectric Nanoantenna. \emph{ACS Photonics} \textbf{2015}, \emph{2},
  1669--1674\relax
\mciteBstWouldAddEndPuncttrue
\mciteSetBstMidEndSepPunct{\mcitedefaultmidpunct}
{\mcitedefaultendpunct}{\mcitedefaultseppunct}\relax
\EndOfBibitem
\bibitem[Kumar \latin{et~al.}(2012)Kumar, Lu, Huck, and Andersen]{FIBbrg}
Kumar,~S.; Lu,~Y.-W.; Huck,~A.; Andersen,~U.~L. Propagation of plasmons in
  designed single crystalline silver nanostructures. \emph{Opt. Express}
  \textbf{2012}, \emph{20}, 24614--24622\relax
\mciteBstWouldAddEndPuncttrue
\mciteSetBstMidEndSepPunct{\mcitedefaultmidpunct}
{\mcitedefaultendpunct}{\mcitedefaultseppunct}\relax
\EndOfBibitem
\bibitem[Park \latin{et~al.}(2012)Park, Ambwani, Manno, Lindquist, Nagpal, Oh,
  Leighton, and Norris]{SingleCrystallineSilverFilm}
Park,~J.~H.; Ambwani,~P.; Manno,~M.; Lindquist,~N.~C.; Nagpal,~P.; Oh,~S.-H.;
  Leighton,~C.; Norris,~D.~J. Single-Crystalline Silver Films for Plasmonics.
  \emph{Advanced Materials} \textbf{2012}, \emph{24}, 3988--3992\relax
\mciteBstWouldAddEndPuncttrue
\mciteSetBstMidEndSepPunct{\mcitedefaultmidpunct}
{\mcitedefaultendpunct}{\mcitedefaultseppunct}\relax
\EndOfBibitem
\bibitem[Drexhage(1970)]{Drexhage}
Drexhage,~K. Influence of a dielectric interface on fluorescence decay time.
  \emph{Journal of Luminescence} \textbf{1970}, \emph{1}, 693 -- 701\relax
\mciteBstWouldAddEndPuncttrue
\mciteSetBstMidEndSepPunct{\mcitedefaultmidpunct}
{\mcitedefaultendpunct}{\mcitedefaultseppunct}\relax
\EndOfBibitem
\bibitem[Ford and Weber(1984)Ford, and Weber]{FordAndWeber}
Ford,~G.; Weber,~W. Electromagnetic interactions of molecules with metal
  surfaces. \emph{Physics Reports} \textbf{1984}, \emph{113}, 195 -- 287\relax
\mciteBstWouldAddEndPuncttrue
\mciteSetBstMidEndSepPunct{\mcitedefaultmidpunct}
{\mcitedefaultendpunct}{\mcitedefaultseppunct}\relax
\EndOfBibitem
\bibitem[Chance \latin{et~al.}(1978)Chance, Proc, and Silbey]{Chance}
Chance,~R.~R.; Proc,~A.; Silbey,~R. Molecular fluorescence and energy transfer
  near interfaces. \emph{Adv. Chem. Phys} \textbf{1978}, \emph{37}, 1 --
  34\relax
\mciteBstWouldAddEndPuncttrue
\mciteSetBstMidEndSepPunct{\mcitedefaultmidpunct}
{\mcitedefaultendpunct}{\mcitedefaultseppunct}\relax
\EndOfBibitem
\bibitem[Avrutsky \latin{et~al.}(2010)Avrutsky, Soref, and
  Buchwald]{High_Low_Index_SPPTheory}
Avrutsky,~I.; Soref,~R.; Buchwald,~W. Sub-wavelength plasmonic modes in a
  conductor-gap-dielectric system with a nanoscale gap. \emph{Opt. Express}
  \textbf{2010}, \emph{18}, 348--363\relax
\mciteBstWouldAddEndPuncttrue
\mciteSetBstMidEndSepPunct{\mcitedefaultmidpunct}
{\mcitedefaultendpunct}{\mcitedefaultseppunct}\relax
\EndOfBibitem
\bibitem[Cui \latin{et~al.}(2015)Cui, Zhang, Liu, Lee, Bracher, Ohno,
  Awschalom, and Hu]{EvelynHu_NVCavity}
Cui,~S.; Zhang,~X.; Liu,~T.-l.; Lee,~J.; Bracher,~D.; Ohno,~K.; Awschalom,~D.;
  Hu,~E.~L. Hybrid Plasmonic Photonic Crystal Cavity for Enhancing Emission
  from near-Surface Nitrogen Vacancy Centers in Diamond. \emph{ACS Photonics}
  \textbf{2015}, \emph{2}, 465--469\relax
\mciteBstWouldAddEndPuncttrue
\mciteSetBstMidEndSepPunct{\mcitedefaultmidpunct}
{\mcitedefaultendpunct}{\mcitedefaultseppunct}\relax
\EndOfBibitem
\bibitem[Yang \latin{et~al.}(2017)Yang, Miller, Christensen, Joannopoulos, and
  Solja{\v c}i{\' c}]{LowLossPlasmonics}
Yang,~Y.; Miller,~O.~D.; Christensen,~T.; Joannopoulos,~J.~D.; Solja{\v c}i{\'
  c},~M. Low-Loss Plasmonic Dielectric Nanoresonators. \emph{Nano Letters}
  \textbf{2017}, \emph{17}, 3238--3245\relax
\mciteBstWouldAddEndPuncttrue
\mciteSetBstMidEndSepPunct{\mcitedefaultmidpunct}
{\mcitedefaultendpunct}{\mcitedefaultseppunct}\relax
\EndOfBibitem
\bibitem[Schell \latin{et~al.}(2011)Schell, Kewes, Schr{\" o}der, Wolters,
  Aichele, and Benson]{PickAndPlace}
Schell,~A.~W.; Kewes,~G.; Schr{\" o}der,~T.; Wolters,~J.; Aichele,~T.;
  Benson,~O. A scanning probe-based pick-and-place procedure for assembly of
  integrated quantum optical hybrid devices. \emph{Review of Scientific
  Instruments} \textbf{2011}, \emph{82}, 073709\relax
\mciteBstWouldAddEndPuncttrue
\mciteSetBstMidEndSepPunct{\mcitedefaultmidpunct}
{\mcitedefaultendpunct}{\mcitedefaultseppunct}\relax
\EndOfBibitem
\bibitem[Jiang \latin{et~al.}(2003)Jiang, Zhang, and
  Zhao]{MeltingPointDepression}
Jiang,~Q.; Zhang,~S.; Zhao,~M. Size-dependent melting point of noble metals.
  \emph{Materials Chemistry and Physics} \textbf{2003}, \emph{82}, 225 --
  227\relax
\mciteBstWouldAddEndPuncttrue
\mciteSetBstMidEndSepPunct{\mcitedefaultmidpunct}
{\mcitedefaultendpunct}{\mcitedefaultseppunct}\relax
\EndOfBibitem
\bibitem[Kurvits \latin{et~al.}(2015)Kurvits, Jiang, and
  Zia]{FourierMicroscopy}
Kurvits,~J.~A.; Jiang,~M.; Zia,~R. Comparative analysis of imaging
  configurations and objectives for Fourier microscopy. \emph{J. Opt. Soc. Am.
  A} \textbf{2015}, \emph{32}, 2082--2092\relax
\mciteBstWouldAddEndPuncttrue
\mciteSetBstMidEndSepPunct{\mcitedefaultmidpunct}
{\mcitedefaultendpunct}{\mcitedefaultseppunct}\relax
\EndOfBibitem
\bibitem[Neu \latin{et~al.}(2011)Neu, Steinmetz, Riedrich-M{\" o}ller, Gsell,
  Fischer, Schreck, and Becher]{SiV}
Neu,~E.; Steinmetz,~D.; Riedrich-M{\" o}ller,~J.; Gsell,~S.; Fischer,~M.;
  Schreck,~M.; Becher,~C. Single photon emission from silicon-vacancy colour
  centres in chemical vapour deposition nano-diamonds on iridium. \emph{New
  Journal of Physics} \textbf{2011}, \emph{13}, 025012\relax
\mciteBstWouldAddEndPuncttrue
\mciteSetBstMidEndSepPunct{\mcitedefaultmidpunct}
{\mcitedefaultendpunct}{\mcitedefaultseppunct}\relax
\EndOfBibitem
\bibitem[Iwasaki \latin{et~al.}(2015)Iwasaki, Ishibashi, Miyamoto, Doi,
  Kobayashi, Miyazaki, Tahara, Jahnke, Rogers, Naydenov, Jelezko, Yamasaki,
  Nagamachi, Inubushi, Mizuochi, and Hatano]{GeV}
Iwasaki,~T. \latin{et~al.}  Germanium-Vacancy Single Color Centers in Diamond.
  \emph{Sc. Rep} \textbf{2015}, \emph{5}, 12882\relax
\mciteBstWouldAddEndPuncttrue
\mciteSetBstMidEndSepPunct{\mcitedefaultmidpunct}
{\mcitedefaultendpunct}{\mcitedefaultseppunct}\relax
\EndOfBibitem
\bibitem[Manfrinato \latin{et~al.}(2013)Manfrinato, Wanger, Strasfeld, Han,
  Marsili, Arrieta, Mentzel, Bawendi, and Berggren]{HolePLacement}
Manfrinato,~V.~R.; Wanger,~D.~D.; Strasfeld,~D.~B.; Han,~H.-S.; Marsili,~F.;
  Arrieta,~J.~P.; Mentzel,~T.~S.; Bawendi,~M.~G.; Berggren,~K.~K. Controlled
  placement of colloidal quantum dots in sub-15 nm clusters.
  \emph{Nanotechnology} \textbf{2013}, \emph{24}, 125302\relax
\mciteBstWouldAddEndPuncttrue
\mciteSetBstMidEndSepPunct{\mcitedefaultmidpunct}
{\mcitedefaultendpunct}{\mcitedefaultseppunct}\relax
\EndOfBibitem
\bibitem[Jiang \latin{et~al.}(2015)Jiang, Kurvits, Lu, Nurmikko, and
  Zia]{ElectrostaticAssembly}
Jiang,~M.; Kurvits,~J.~A.; Lu,~Y.; Nurmikko,~A.~V.; Zia,~R. Reusable Inorganic
  Templates for Electrostatic Self-Assembly of Individual Quantum Dots,
  Nanodiamonds, and Lanthanide-Doped Nanoparticles. \emph{Nano Letters}
  \textbf{2015}, \emph{15}, 5010--5016\relax
\mciteBstWouldAddEndPuncttrue
\mciteSetBstMidEndSepPunct{\mcitedefaultmidpunct}
{\mcitedefaultendpunct}{\mcitedefaultseppunct}\relax
\EndOfBibitem
\bibitem[Ramachandran \latin{et~al.}(2009)Ramachandran, Kristensen, and
  Yan]{FiberModeConv}
Ramachandran,~S.; Kristensen,~P.; Yan,~M.~F. Generation and propagation of
  radially polarized beams in optical fibers. \emph{Opt. Lett.} \textbf{2009},
  \emph{34}, 2525--2527\relax
\mciteBstWouldAddEndPuncttrue
\mciteSetBstMidEndSepPunct{\mcitedefaultmidpunct}
{\mcitedefaultendpunct}{\mcitedefaultseppunct}\relax
\EndOfBibitem
\bibitem[Johnson and Christy(1972)Johnson, and Christy]{JohnsonChristy}
Johnson,~P.~B.; Christy,~R.~W. Optical Constants of the Noble Metals.
  \emph{Phys. Rev. B} \textbf{1972}, \emph{6}, 4370--4379\relax
\mciteBstWouldAddEndPuncttrue
\mciteSetBstMidEndSepPunct{\mcitedefaultmidpunct}
{\mcitedefaultendpunct}{\mcitedefaultseppunct}\relax
\EndOfBibitem
\end{mcitethebibliography}
\end{document}